# Structure of nanoscale-pitch helical phases: blue phase and twist-bend nematic phase resolved by resonant soft X-ray scattering

*Mirosław Salamończyk[§†], Nataša Vaupotič[‡∞*], Damian Pociecha[∥], Cheng Wang[§], Chenhui Zhu[§*], Ewa Gorecka[∥*]*

[§]Advance Light Source, Lawrence Berkeley National Laboratory, Berkeley, CA 94720, USA

[†]Liquid Crystal Institute & Department of Physics, Kent State University, Kent, OH 44242, USA

[‡]Department of Physics, Faculty of Natural Sciences and Mathematics, University of Maribor, Koroška 160, 2000 Maribor, Slovenia

[∞]Jozef Stefan Institute, Jamova 39, 1000 Lubljana, Slovenia

[∥]Faculty of Chemistry, University of Warsaw, Zwirki i Wigury 101, 02-089 Warsaw, Poland

Periodic structures of phases with orientational order of molecules, but homogenous electron density distribution: a short pitch cholesteric, blue phase and twist-bend nematic phase, were probed by a resonant soft x-ray scattering (RSoXS) at the carbon K-edge. The theoretical model shows that in case of a simple heliconical nematic structure two resonant signals corresponding to the full and half pitch band should be present, while only the full pitch band is observed in experiment. This suggests that the twist-bend nematic phase has complex structure with a double-helix, built of two interlocked, shifted helices. We confirm that the helical pitch in the twist-bend nematic phase is in a 10 nm range, for both, the chiral and achiral materials. We also show that the symmetry of a blue phase can unambiguously be determined through a resonant enhancement of x-ray diffraction signals, by including polarization effects, which are found to be an important indicator in phase structure determination.

Nanostructured soft materials with hierarchical organization have recently attracted a lot of attention in both fundamental research and applications, the examples of such systems can be found among liquid crystals, gels or structured polymers. The advantage of chemical diversity of such materials, ranging from low weight molecules to polymers and nanoparticles, is that it enables for intelligent material design by tuning of particular material functional property, as electronic or optical energy band gap, polar or magnetic order, etc. Some of these materials may exhibit helical structures or helical morphology, the 'classical' examples of such systems are DNA and peptides, in which helical arrangement of molecules is induced by chirality of molecular building blocks. The verity of helical structures formed by low weight molecules can be found among liquid crystals;[1] chiral mesogenic molecules can self-assemble into a helical nematic or smectic phases, in which molecules uniformly twist, or into a blue phases, a three-dimensional, nanoscale cubic phases formed of double twist cylinders, where each cylinder has an internal, nanoscale helical structure.[1] Less frequent are examples of achiral mesogens (bent-core, dimers, etc.) that exhibit nanoscale helical structures[2]: filaments made either from soft crystal layers[3-5] or smectic membranes,[6] as well as a helioconical nematic phase with an ultra-short, nano-scale helical pitch.[7-14] These materials have attracted recently lot of attention due to the rich physical phenomena related to spontaneous symmetry breaking. However, the origin of the helix formation in achiral martials is still on debate, partially due to limited number of in-situ, experimental probes of orientational order at submicron scale. For some soft matter phases the orientational order of molecules is coupled to density modulations, and therefore their structure can be revealed by standard x-ray diffraction technique, but the phases with a uniform electron density, such as nematic, or blue phases cannot be distinguished by this technique. The method that is sensitive to the spatial variation of the orientation of molecules at non-scale is a resonant x-ray scattering, which has so far mainly been operated at the absorption edges of Bromine, Selenium and Sulfur for studying smectic C liquid crystal subphases, to determine their secondary structure induced by the molecular orientation.[15-18] Recently the resonant soft X-ray scattering at the carbon absorption edge has been applied to study a phase separation in block copolymers,[19,20] molecular orientation in solar cells,[21] and morphology of helical nanofilaments (B$_4$ phase) made of bent-core mesogens.[3]

Here we demonstrate that the RSoXS in combination with theoretical modelling can be used to reveal a 3D structure of phases with a uniform electron density and a periodically modulated orientational order. Furthermore, we show that the polarization analysis is very important in removing the remaining structural degeneracy. We apply the method to blue phases (BP),[22] chiral nematic (N$^*$) and twist-bend nematic (N$_{TB}$) phases. The structure of all these phases (Fig. 1) is helical and the twist originates either from the molecular chirality (N* and BP phases) or is caused by a unique, bent molecular geometry (N$_{TB}$ phase).

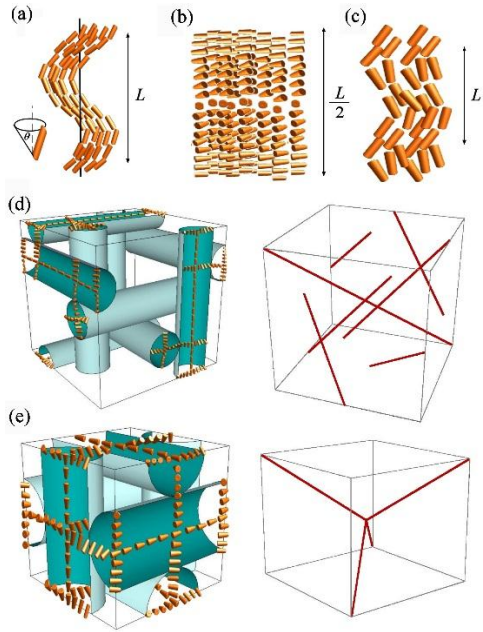

Figure 1. Structure of the (a) twist-bend nematic N$_{TB}$, (b) chiral nematic N*, (c) splay-bend nematic N$_{SB}$. $L$ is a modulation length. Double twisted cylinders (left) and the defect network (right) in blue phases (d) of type I (BPI) and (e) type II (BPII).

The N* phase has a single twisted helical structure, with the average direction of long molecular axes (director) lying perpendicular to the helix axis. The generally accepted model for the N$_{TB}$ phase is a single, uniform helix with the director processing around the helix axis at some angle $\theta$. For the bent molecules, the non-helical splay-bend nematic (N$_{SB}$) phase was also predicted,[23,24] in which the modulation of the long molecular axis direction is constrained to the plane. In blue phases the twist is induced in every direction perpendicular to the director, resulting in the so called double twist (DT) cylinder. Such DT cylinders cannot continuously fill the space, thus a 3D network of defects is formed with either the body centered (BPI, $I4_132$ symmetry) or simple cubic (BPII, $P4_232$ symmetry) structure. By the elastic x-ray scattering only diffused signals related to the short range positional order are detected in all these phases. For the BP or N$^*$ phases having the helical pitch in the visible or near IR wavelength range, optical methods are used to determine the structure parameters. Here we provide a direct, effective and general approach that can be applied also to structures with periodicities below the optical wavelength, to which neither optical nor classical x-ray diffraction techniques are sensitive to. By RSoXS, the information about the molecular orientational order with the periodicities of the order of a few to hundreds of nanometers can easily be obtained.

We have performed RSoXS measurements for a chiral **SB3**[14] material showing a N$^*$ - N$_{TB}$ phase sequence and a chiral **AZO7**[14,25] compound showing a BP - N$_{TB}$ phase sequence on cooling and N$_{TB}$ - N$^*$ - BP phase sequence on heating (Fig. 2). The results were compared to those obtained for the achiral **CB7CB** dimeric compound, which is a model N$_{TB}$ material showing a N - N$_{TB}$ phase sequence.[8-11,13]

When the chiral material, **AZO7**, is cooled from the isotropic phase, several RSoXS signals are obtained, that can be indexed to the cubic structure (Fig. 3). On further cooling, the signals related to the cubic lattice disappear and a single peak corresponding to a much shorter periodicity appears, signifying a transition to the N$_{TB}$ phase. The position of the signal in the N$_{TB}$ phase is temperature dependent: on cooling the periodicity reduces from 20.3 to 13.3 nm (see SI, Fig. S15). Interestingly, on a subsequent heating, the sample undergoes a transition from the N$_{TB}$ to the N$^*$ phase, in which only one signal corresponding to the half pitch band is observed, at 110 nm (Fig. 3). On further heating, the BP phase is observed only in a narrow temperature range of few degrees, close to the isotropic phase.

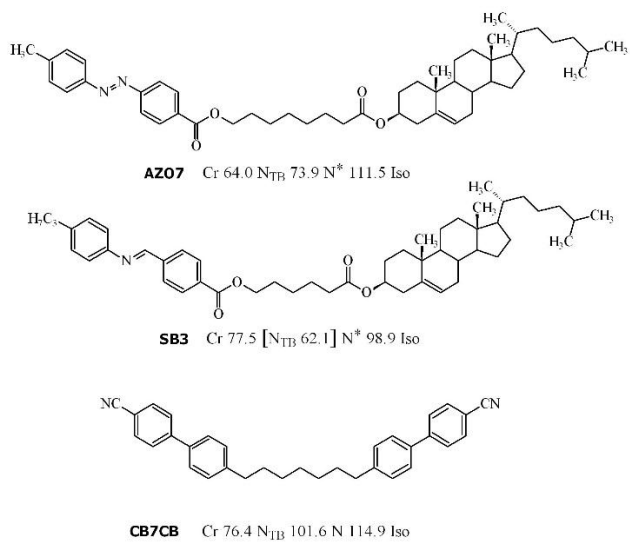

**Figure 2.** Molecular structure of the studied compounds, **AZO7**, **SB3** and **CB7CB**. For each compound a phase sequence and phase transition temperatures (°C) determined by the differential scanning calorimetry (DSC) on heating scans are given. Note, that for **AZO7** a narrow range of a blue phase between the N* and Iso has been found by microscopic observations; however, it was not recorded on the DSC curves due to a limited resolution. Upon cooling the **AZO7** samples, the blue phase was metastable down to the transition to the $N_{TB}$ phase and thus the cholesteric phase was not observed.

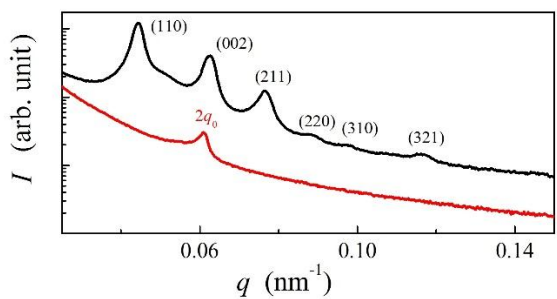

**Figure 3**. RSoXS patterns, the intensity ($I$) in arbitrary units as a function of the magnitude of the scattering vector $q$ for the **AZO7** compound in the BPI (black line) and N* (red line).

Apparently, for this material, the BP phase is thermodynamically stable close to the clearing temperature but can easily be supercooled, similarly as observed previously for other dimeric materials.[26] Such a phase sequence enables a direct comparison between the cholesteric pitch and the size of the unit cell of the BP phase. Assuming that the helical pitch at the transition from the cholesteric to the BP phase does not change significantly, the unit cell size in the BPI phase ($I4_132$ symmetry) should correspond to the full pitch length ($L$) in $N^*$ and to the half pitch length ($L/2$) in the BPII phase ($P4_232$ symmetry) (see Fig. 1). Because the first signal in the BP phase of **AZO7** appears at a lower value of $q$ than in the $N^*$ phase, the observed blue phase must have the $I4_132$ symmetry (BPI). From the position of the main signals observed in the RSoXS pattern of the cubic phase (peaks (*110*), (*002*) and (*211*)) the crystallographic lattice parameter (*a*) is obtained: $a = 201$ nm. Except for the (*002*) peak the observed peaks are allowed by the symmetry, they are thus resonantly enhanced. A comparison of the position of the purely resonant (*002*) signal in the BP phase and the half pitch band signal in the $N^*$ shows that, within the experimental error, the pitch in the BP phase does not change at the phase transition. In both the BPI and $N^*$ phase, the azimuthal position of the signals was strongly sensitive to the polarization of the incoming beam (Fig. 4b,c).

For the **SB3** material, on cooling, the resonant peak corresponding to the half pitch band ($L/2 \approx$ 112 nm) is detected in the $N^*$ phase temperature range (Fig. S10). Upon the transition to the $N_{TB}$ phase, the signal corresponding to 11 nm develops. Interestingly, the periodicity detected by the RSoXS experiment in the $N_{TB}$ phase is much smaller than the one measured by the atomic force microscopy (AFM) method, where the fingerprint texture with lines separated by 50-80 nm was observed.[25] The short pitch periodic structure detected by the resonant x-ray method is in line with a large compressibility modulus measured previously.[25] Despite many efforts, the long periodicity detected by the AFM measurements (see SI, Fig. S13), was not observed by the RSoXS method.

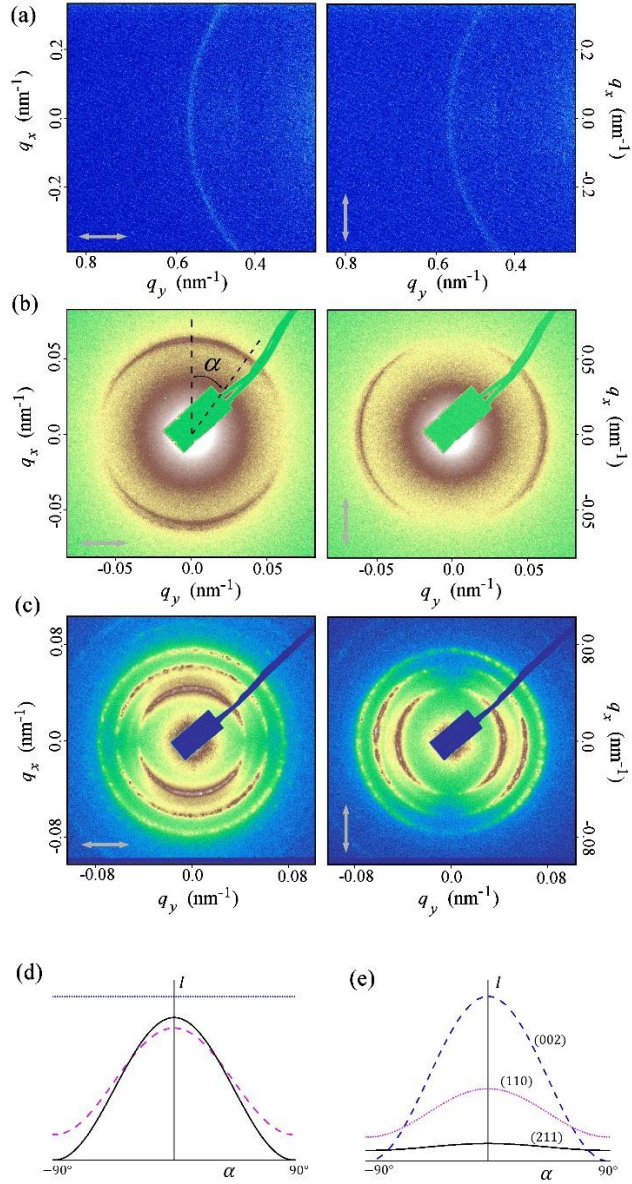

**Figure 4.** RSoXS patterns of the **AZO7** compound in the (a) $N_{TB}$, (b) $N^*$ and (c) BPI phases recorded for two perpendicular polarizations of the incident beam. d) Theoretically calculated intensity in arbitrary units as a function of the azimuthal angle $\alpha$; $2q_0$ peak in $N^*$ (black solid line); $2q_0$ (magenta dashed line) and $q_0$ (blue dotted line) peak in $N_{TB}$, both calculated at the cone angle $\theta = 10$ deg. To present the results on the same graph, the intensity of the $N^*$ peak was divided by 10 and the intensity of the $2q_0$ peak in $N_{TB}$ was multiplied by 100. (e) The intensities of the three most intensive peaks in the BPI phase. Parameter values: $f_1/f_2 = 0.6$, where $f_1$, $f_2$ and $-(f_1+f_2)$ are the eigenvalues of the local traceless dispersion correction to the form factor (see SI).

It is not obvious at the moment, what caused the large periodicity structure of the phase detected by the AFM, but one possibility is that the surface free samples, studied by the AFM, have an additional structure induced by chirality, and surfaces cause its destabilization when a thin cell (~ 0.5 μm thick) is used in the x-ray studies reported here.The RSoXS measurements were also repeated for **CB7CB**, the model $N_{TB}$ material.[27] For this compound the resonant peak was also found in the crystal phase: the signal corresponding to the periodicity of 7.8 nm gives the full size of the crystallographic unit cell and corresponds roughly to three molecular distances. Such a signal is forbidden for the $P3_121$ symmetry (the crystal structure was resolved for the **CB9CB** homologue[28]) and is thus not observed in the elastic diffraction experiments. The melting of the crystal and the transition to the $N_{TB}$ phase is associated with a sudden jump of the peak position to 8.2 nm (Fig. 5). This clearly shows that the structures of the crystal and $N_{TB}$ phases are closely related and probably caused by steric interactions[24,29] rather than by the flexoelectric effect.[23,30,31]

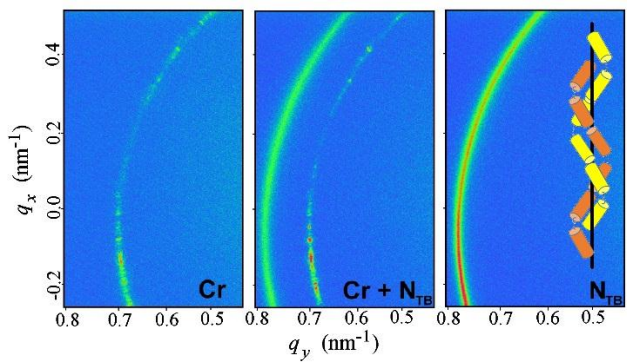

**Figure 5**. Two dimensional RSoXS patterns for the **CB7CB** compound in the crystalline (Cr) and twist-bend nematic phase ($N_{TB}$) and in the temperature range of the phase coexistence. The observed signals correspond to the periodicities of 7.8 nm in Cr and 8.2 nm in $N_{TB}$. In the inset – a simplified model of $N_{TB}$ phase structure with two interlocked, mutually shifted helices.

Interestingly, scattering patterns of N* and BP show strong intensity anisotropy, but that of the N$_{TB}$ does not. To fully understand the resonant x-ray diffraction pattern, one has to calculate the dispersion correction to the structure factor, which is proportional to the Fourier transform of the spatially dependent structure polarizability.[32,33] We have obtained the dispersion correction for the LC phases shown in Fig. 1 and found the scattered intensity as a function of the polarization direction of the incident beam and biaxiality of the x-ray susceptibility. Details of calculations are given in the Supporting Information, here we give only the main results.

In N* and N$_{SB}$ phases only the half pitch peak at $2q_0$ is expected, where $q_0=2\pi/L$ is the magnitude of the wave vector of the helical structure and $L$ is the helical pitch. In N* the $2q_0$ signal intensity is very sensitive to the polarization of the incident x-ray beam, as expected, because the beam with the electric field polarized perpendicular to the helix detects the highest contrast in the director modulation, while the beam polarized along the helix axis detects only the uniform structure of the short molecular axes. As a result, for a powder sample, in which the helix axis directions are randomly distributed in space, a 2D scattering pattern exhibits anisotropy for the linearly polarized incident x-ray beam. Two complementary parts of the diffraction ring are predicted in the diffraction pattern, if the polarization of the incident beam is rotated by 90 deg. For the powder sample of the N$_{SB}$ phase the signal is almost insensitive to the polarization and is also independent of the magnitude of the modulation angle. For the N$_{TB}$ phase with the heliconical spatial variation of the director, the model predicts two signals, at $q_0$ and $2q_0$. Interestingly, in the N$_{TB}$ phase the $q_0$ and $2q_0$ signals have a different sensitivity to the polarization of the x-ray beam; only for the $2q_0$ signal the dependence is similar to the one observed in the N* phase, while the $q_0$ signal is invariant to the polarization direction (SI, Fig. S3). Experimentally, only one peak has been found in the N$_{TB}$ phase, which shows no intensity anisotropy, suggesting that the peak is the $q_0$ signal, corresponding to a 360 deg rotation of molecular director around the helical axis. It should be pointed out that for none of the $N_{TB}$ materials studied so far the signal corresponding to a half pitch band ($2q_0$) was detected. Previous RSoXS experiments at the carbon K-edge on **CB7CB**[27] as well as measurements performed using the x-ray scattering at the Selenium K-edge,[34] also give a systematic lack of the $2q_0$ signal in the $N_{TB}$ phase. A possible explanation is that the structure is actually made of two interlocked helices, which are mutually shifted (Fig. 5). The calculations show that the $2q_0$ peak is canceled out for the shift between helices equal to $L/4$. The shift can be induced by short-range interactions of dimers that favor a local intercalated arrangement of the neighboring molecules,[29] which seems to be quite common for dimeric molecules.[34] The tendency for the formation of interlocked helices is probably general for bent dimers, as the crystal structure of the **CB9CB** homologue[28] is also formed of interlocked and shifted helices, which are made of two types of slightly different molecular conformers.

For the blue phases formed of double twisted cylinders, we have calculated the positions of the elastic diffraction peaks by building up the structure from infinitely long cylinders. The Fourier transform of the cubic lattice filled with cylinders of a finite width and uniform density formally gives the ($hkl$) diffraction signals if at least one of the Miller index, $h$, $k$ or $l$, is zero (SI, Tab. S1). However, because the cylinders are in contact, the density of the unit cell is practically uniform and none of these peaks is observed in the elastic scattering. If the electron density in the defect regions between the cylinders is different than in cylinders (Fig. 1) the signals allowed by the $P4_232$ (BPII) or $I4_132$ (BPI) symmetry should be detected. In practice, none of these signals is found by the elastic x-ray scattering, confirming that the electron density difference between the cylinders and defect regions is negligible. In order to include the resonant effects, the dispersion correction was calculated due to the helical spatial variation of the director in the direction perpendicular to the cylinder axis and by arranging the cylinders into a proper cubic lattice (Fig. 1). We show that some signals allowed by the symmetry of the BP phases, but not detectable by the elastic x-ray diffraction, become visible in the resonant x-ray diffraction due to their enhancement by orientation modulations (Tab. 1). Additionally, purely resonant signals: (*001*) for *P*4$_2$32 and (*002*) for *I*4$_1$32 are predicted.

**Table 1**. The resonant or resonantly enhanced diffraction signals in the BPI and BPII phases. Ticks denote the peaks with a non-zero intensity. The (*002*) peak in the BPI, and the (*001*) peak in the BPII are purely resonant signals; the other observable signals are resonantly enhanced.

| (*hkl*) | (*001*) | (*110*) | (*111*) | (*002*) | (*210*) | (*211*) | (*220*) |
|---------|---------|---------|---------|---------|---------|---------|---------|
| BPII | ✓ | ✓ | ✓ | ✓ | ✓ | ✓ | ✓ |
| BPI | x | ✓ | x | ✓ | x | ✓ | ✓ |

| (*hkl*) | (*221*) | (*310*) | (*311*) | (*222*) | (*320*) | (*321*) |
|---------|---------|---------|---------|---------|---------|---------|
| BPII | ✓ | ✓ | ✓ | x | ✓ | ✓ |
| BPI | x | x | x | ✓ | x | ✓ |

In the BPI phase intensities of all the signals are sensitive to the polarization of the incident x-ray beam. For a powder sample, the two most intensive signals, (*110*) and (*002*), have the same polarization dependence as the half-pitch band signal in the N$^*$ phase and the complementary parts of the diffraction rings are predicted for the incident beams polarized in perpendicular directions. The polarization dependence of the (*211*) signal is defined by the biaxiality of the susceptibility.

For strongly biaxial molecules, such as bent dimers, the dependence is similar to that observed for the (*110*) and (*002*) signals.

Summarizing, we have shown that the comparison between the RSoXS signals in blue phases and N* phase enable a straightforward determination of the type of the BP. We have also shown that the modulation pitch in the $N_{TB}$ phase is of the same order of magnitude for both, the chiral and achiral materials. However, the structure of the twist-bend nematic phase turns to be more complex than commonly accepted. The theoretical model shows that in the case of the heliconical structure RSoXS signals corresponding to the full and half pitch band should both be present and they should have a very different polarization dependence. Experimentally, only one signal was found, with the intensity independent of the beam polarization; it was thus unambiguously identified as the full pitch band. The lack of the half pitch band strongly suggests that the $N_{TB}$ structure is made of two interlocked and shifted helices.

ASSOCIATED CONTENT

**Supporting Information**. Details of the theoretical model, materials, experimental methods and additional results.

AUTHOR INFORMATION

**Corresponding Author**

* N. Vaupotič: natasa.vaupotic@um.si

* C. Zhu: chenhuizhu@lbl.gov

* E. Gorecka : gorecka@chem.uw.edu.pl

**Author Contributions**

The manuscript was written through contributions of all authors. All authors have given approval to the final version of the manuscript.

ACKNOWLEDGMENT

MS acknowledges the support of the U.S. National Science Foundation I2CAM International Materials Institute Award, Grant DMR-1411344 and NSF grant DMR-1307674.

NV acknowledges the financial support from the Slovenian Research Agency (research core funding No. P1-0055). EG acknowledges the support of the National Science Centre (Poland) under the grant no. UMO-2015/19/P/ST5/03813.

The authors acknowledge the important technical help and discussions from Drs. A. Kilcoyne, M. Brady, G. Su and A. Hexemer at the ALS LBNL and the synthesis of the studied materials by A. Zep.

The beam line 11.0.1.2 at the Advanced Light Source is supported by the Director of the Office of Science, Office of Basic Energy Sciences, of the U.S. Department of Energy under Contract No. DE-AC02- 05CH11231.

REFERENCES

[1] *Chirality in Liquid Crystals*, ed. Kitzerow, H.; Bahr, C., Springer-Verlag New York, **2001**.

[2] Tschierske, C.; Ungar G. Mirror Symmetry Breaking by Chirality Synchronisation in Liquids and Liquid Crystals of Achiral Molecules, *ChemPhysChem,* **2016**, 17, 9 – 26

[3] Zhu, C.; Wang, C.; Young, A.; Liu, F.; Gunkel, I.; Chen, D.; Walba, D.; Maclennan, J.; Clark, N.; Hexemer, A. Probing and Controlling Liquid Crystal Helical Nanofilaments. *Nano Lett*. **2015**, *15*, 3420–3424.

[4] Hough, L. E.; Jung, H. T.; Krüerke, D.; Heberling, M. S.; Nakata, M.; Jones, C. D.; Chen, D.; Link, D. R.; Zasadzinski, J.; Heppke, G.; Rabe, J. P.; Stocker, W.; Körblova, E.; Walba, D. M.; Glaser, M. A.; Clark, N. A. Helical Nanofilament Phases, Science 2009, 325, 456-460.

[5] Matraszek, J.; Topnani, N.; Vaupotič, N.; Takezoe, H.; Mieczkowski, J.; Pociecha, D.; Gorecka, E. Monolayer Filaments versus Multilayer Stacking of Bent-Core Molecules, *Angew. Chem. Int. Ed.* **2016**, *55*, 3468–3472.

[6] Hough, L. E.; Spannuth, M.; Nakata, M.;. Coleman, D. A.; Jones, C. D.; Dantlgraber, G; Clark, N. A. Chiral Isotropic Liquids from Achiral Molecules. *Science* **2009**, *325*, 452-456.

[7] Pelzl G.; Eremin A.; Diele S.; Kresse H.; Weisflog W. Spontaneous chiral ordering in the nematic phase of an achiral banana-shaped compound. *J. Mater. Chem*. **2002,** *12*, 2591-2593.

[8] Cestari M.; Diez-Berart S.; Dunmur D. A.; Ferrarini A.; de la Fuente M. R.; Jackson D. J. B.; Lopez D. O.; Luckhurst G. R.; Perez-Jubindo M. A.; Richardson R. M.; Salud J.; Timimi

B. A.; Zimmermann H. Phase behavior and properties of the liquid-crystal dimer 1″,7″-bis(4-cyanobiphenyl-4′-yl) heptane: A twist-bend nematic liquid crystal. *Phys. Rev. E* **2011**, *84*, 031704.

[9] Meyer, C.; Luckhurst, G. R.; Dozov, I. Flexoelectrically Driven Electroclinic Effect in the Twist-Bend Nematic Phase of Achiral Molecules with Bent Shapes. *Phys. Rev. Lett*. **2013**, *111*, 067801.

[10] Chen, D.; Porada, J. H.; Hooper, J. B.; Klittnick, A.; Shen, Y.; Tuchband, M. R.; Korblova, E.; Bedrov, D.; Walba, D. M.; Glaser, M. A.; et al. Chiral Heliconical Ground State of Nanoscale Pitch in a Nematic Liquid Crystal of Achiral Molecular Dimers. *Proc. Natl. Acad. Sci*. **2013**, *110*, 15931–15936.

[11] Borshch, V.; Kim, Y.-K.; Xiang, J.; Gao, M.; Jakli, A.; Panov, V. P.; Vij, J. K.; Imrie, C. T.; Tamba, M. G.; Mehl, G. H.; et al. Nematic Twist-Bend Phase with Nanoscale Modulation of Molecular Orientation. *Nat. Commun*. **2013**, *4*, 1–8.

[12] Chen D.; Nakata M.; Shao R.; Tuchband M. R.; Shuai M.; Baumeister U.; Weissflog W.; Walba D. M.; Glaser M. A.; Maclennan J. E.; Clark N. A. *Phys. Rev. E* **2014**, *89*, 022506.

[13] Górecka, E.; Salamończyk, M.; Zep, A.; Pociecha, D.; Welch, C.; Ahmed, Z.; Mehl, G. H. Do the Short Helices Exist in the Nematic TB Phase? *Liq. Cryst*. **2015**, *42*, 1–7.

[14] Zep, A.; Aya, S.; Aihara, K.; Ema, K.; Pociecha, D.; Madrak, K.; Bernatowicz, P.; Takezoe, H.; Gorecka, E. Multiple Nematic Phases Observed in Chiral Mesogenic Dimers. J. Mater. Chem. C 2013, 1, 46.

[15] Mach, P.; Pindak, R.; Levelut, A.-M.; Barois, P.; Nguyen, H.; Huang, C.; Furenlid, L. Structural Characterization of Various Chiral Smectic- C Phases by Resonant X-Ray Scattering. *Phys. Rev. Lett.* **1998**, *81*, 1015–1018.

[16] Fernandes, P.; Barois, P.; Wang, S. T.; Liu, Z. Q.; McCoy, B. K.; Huang, C. C.; Pindak, R.; Caliebe, W.; Nguyen, H. T. Polarization Studies of Resonant Forbidden Reflections in Liquid Crystals. *Phys. Rev. Lett.* **2007**, *99*, 1–4.

[17] Folcia, C. L.; Ortega, J.; Etxebarria, J.; Rodríguez-Conde, S.; Sanz-Enguita, G.; Geese, K.; Tschierske, C.; Ponsinet, V.; Barois, P.; Pindak, R.; et al. Spontaneous and Field-Induced Mesomorphism of a Silyl-Terminated Bent-Core Liquid Crystal as Determined from Second-Harmonic Generation and Resonant X-Ray Scattering. *Soft Matter* **2014**, *10*, 196–205.

[18] Gleeson, H. F.; Hirst, L. S. Resonant X-Ray Scattering: A Tool for Structure Elucidation in Liquid Crystals. *ChemPhysChem* **2006**, *7*, 321–328.

[19] Wang, C.; Lee, D. H.; Hexemer, A.; Kim, M. I.; Zhao, W.; Hasegawa, H.; Ade, H.; Russell, T. P. Defining the Nanostructured Morphology of Triblock Copolymers Using Resonant Soft X-Ray Scattering. *Nano Lett.* **2011**, *11*, 3906–3911.

[20] Virgili, J. M.; Tao, Y. F.; Kortright, J. B.; Balsara, N. P.; Segalman, R. a. Analysis of Order Formation in Block Copolymer Thin Films Using Resonant Soft X-Ray Scattering. *Macromolecules* **2007**, *40*, 2092–2099.

[21]   Tumbleston, J. R.; Collins, B. a.; Yang, L.; Stuart, A. C.; Gann, E.; Ma, W.; You, W.; Ade, H. The Influence of Molecular Orientation on Organic Bulk Heterojunction Solar Cells. *Nat. Photonics* **2014**, *8*, 385–391.

[22]   Wright, D. C.; Mermin, N. D. Crystalline Liquids: The Blue Phases. *Rev. Mod. Phys.* **1989**, *61*, 385–432.

[23]   Meyer R. B. in Structural Problems in Liquid Crystal Physics, edited by Balian R. and Weil G., Les Houches Summer School in Theoretical Physics, 1973. *Molecular Fluids*, Gordon and Breach, New York, **1976**, 273–373.

[24]   Dozov, I. On the Spontaneous Symmetry Breaking in the Mesophases of Achiral Banana-Shaped Molecules. *Europhys. Lett.* **2007**, *56*, 247–253.

[25]   Gorecka, E.; Vaupotič, N.; Zep, A.; Pociecha, D.; Yoshioka, J.; Yamamoto, J.; Takezoe, H. A Twist-Bend Nematic ($N_{TB}$) Phase of Chiral Materials. *Angew. Chemie Int. Ed.* **2015**, *54*, 10155–10159.

[26]   Coles, H. J.; Pivnenko, M. N. Liquid Crystal "Blue Phases" with a Wide Temperature Range. *Nature* **2005**, *436*, 997–1000.

[27]   Zhu, C.; Tuchband, M. R.; Young, A.; Shuai, M.; Scarbrough, A.; Walba, D. M.; Maclennan, J. E.; Wang, C.; Hexemer, A.; Clark, N. A. Resonant Carbon K-Edge Soft X-Ray Scattering from Lattice-Free Heliconical Molecular Ordering: Soft Dilative Elasticity of the Twist-Bend Liquid Crystal Phase. *Phys. Rev. Lett.* **2016**, *116*, 147803.

[28]   Hori, K.; Iimuro, M.; Nakao, A.; Toriumi, H. Conformational Diversity of Symmetric Dimer Mesogens, A,ω- bis(4,4′-Cyanobiphenyl)octane,    -Nonane,    A,ω-bis(4-

Cyanobiphenyl-4′-yloxycarbonyl)propane, and -Hexane in Crystal Structures. *J. Mol. Struct*. **2004**, *699*, 23–29.

[29]  Vaupotič, N.; Curk, S.; Osipov, M. A.; Čepič, M.; Takezoe, H.; Gorecka, E. Short-Range Smectic Fluctuations and the Flexoelectric Model of Modulated Nematic Liquid Crystals. *Phys. Rev. E* **2016**, *93*, 22704.

[30]  Shamid S. M.; Dhakal S.; Selinger J. V. Statistical mechanics of bend flexoelectricity and the twist-bend phase in bent-core liquid crystals. Phys. Rev. E **2013**, 87, 052503.

[31]  Vaupotič N.; Čepič M.; Osipov M. A.; Gorecka E. Flexoelectricity in chiral nematic liquid crystals as a driving mechanism for the twist-bend and splay-bend modulated phases. Phys. Rev. E **2014**, 89, 030501(R).

[32]  Templeton, D. H.; Templeton, L. K. Polarized X-Ray Absorption and Double Refraction in Vanadyl Bisacetylacetonate. *Acta Crystallogr. Sect. A* **1980**, *36*, 237–241.

[33]  Dmitrienko, V. E. Forbidden Reflections due to Anisotropic X-ray Susceptibility of Crystals. *Acta Crystallogr. Sect. A* **1983**, *39*, 29–35.

[34]  Stevenson, W. D.; Ahmed, Z.; Zeng, X. B.; Welch, C.; Ungar, G.; Mehl, G. H. Molecular Organization in the Twist-Bend Nematic Phase by Resonant X-Ray Scattering at the Se K-Edge and by SAXS , WAXS and GIXRD. *arXiv* **2016**, 1612.01180.

# Structure of nanoscale-pitch helical phases: blue phase and twist-bend nematic phase resolved by resonant soft X-ray scattering

M. Salamończyk, N. Vaupotič, D. Pociecha, C. Wang, C. Zhu, E. Gorecka

## *1.  Supplemental discussion on theoretical modelling*

***Modulated nematic phases:***

Due to the uniform electron density, the elastic x-ray scattering in nematic phases detects only a short range positional order of molecules, while the resonant scattering [1, 2] provides the information on a non-uniform orientational structure of molecules. To calculate the dispersion correction to the form factor we start from the x-ray polarizability of a molecule written in the eigen system of the molecule with the long molecular axis along the $z$-axis. We assume [3], that the form factor in the eigen system ($F_{ei}$) has a form of a traceless tensor (as does the anisotropic part of the polarizability tensor):

$$F_{ei} = \begin{pmatrix} f_1 & 0 & 0 \\ 0 & f_2 & 0 \\ 0 & 0 & -(f_1 + f_2) \end{pmatrix}.$$

In the $N_{TB}$ phase the averaged direction of the long molecular axes (director) is inclined from the $z$ axis of the laboratory frame and rotates along the laboratory $z$ axis (Fig. S1(a)), while in the $N_{SB}$ phase, the director oscillates along the $z$ axis (Fig. S1(b)). The cholesteric phase can be described as a special case of the $N_{TB}$ phase with a conical angle equal to 90 degrees.

The dispersion correction in the laboratory system is obtained by the rotation of the molecular coordinate system. In the $N_{TB}$ phase we first rotate it by an angle $\theta_{TB}$ (the conical angle) around the $y$ axis and then by an angle $\varphi = q_{TB}z$ around the $z$ axis. The dispersion correction tensor for the $N_{TB}$ phase is thus

$$F^{(TB)} = R_\varphi^T \left( R_\theta^T F_{ei} R_\theta \right) R_\varphi \ ,$$

where $R_\theta$ is:

$$R_\theta = \begin{pmatrix} \cos\theta_{TB} & 0 & \sin\theta_{TB} \\ 0 & 1 & 0 \\ -\sin\theta_{TB} & 0 & \cos\theta_{TB} \end{pmatrix}$$

and $R_\varphi$ is

$$R_\varphi = \begin{pmatrix} \cos\varphi & \sin\varphi & 0 \\ -\sin\varphi & \cos\varphi & 0 \\ 0 & 0 & 1 \end{pmatrix} .$$

Calculating the Fourier transform of $F^{(TB)}$ ($F_q^{(TB)}$), we find that the elements of the dispersion correction tensor are different from zero only when $q_z$ has one of the following values: $0$, $\pm q_0$ or $\pm 2q_0$, where $q_0 = 2\pi/L$ is the magnitude of the wave vector of the heliconical deformation with the pitch $L$. If $q_z = q_0$, the dispersion correction to the form factor is:

$$F_{q_0}^{(TB)} = \left(f_1 + \frac{f_2}{2}\right) \sin(2\theta_{TB}) \begin{pmatrix} 0 & 0 & 1 \\ 0 & 0 & i \\ 1 & i & 0 \end{pmatrix}$$

and if $q_z = 2q_0$:

$$F_{2q_0}^{(TB)} = \left[\frac{1}{2}(f_1 - f_2) + \left(f_1 + \frac{f_2}{2}\right)\sin^2\theta_{TB}\right] \begin{pmatrix} 1 & i & 0 \\ i & -1 & 0 \\ 0 & 0 & 0 \end{pmatrix} .$$

In order to find the dispersion correction tensor in the $N_{SB}$ phase we repeat the above procedure, but in this case we have to rotate the director only around the $y$ axis by an angle $\theta = \theta_{SB} \sin q_0 z$, where $\theta_{SB}$ is the magnitude of the modulation angle and $q_0$ is the modulation wave vector. In the $N_{SB}$ phase we find the resonant peak with $q_z = 2q_0$ only:

$$F_{2q_0}^{(SB)} = \left(f_1 + \frac{f_2}{2}\right) \begin{pmatrix} 1 & 0 & i \\ 0 & 0 & 0 \\ i & 0 & -1 \end{pmatrix} .$$

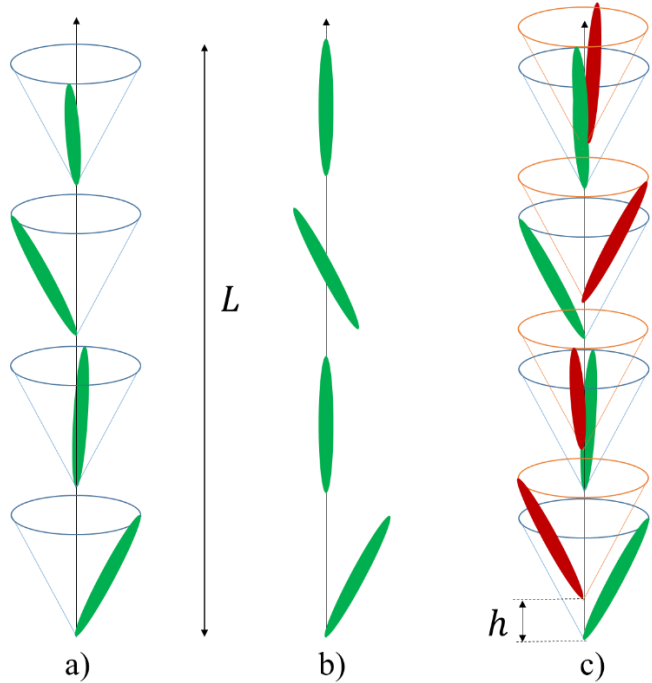

**Fig. S1.** Schematic presentation of the a) N$_{TB}$ and b) N$_{SB}$ phase. c) Modulated nematic phase with two helices with molecules on the opposite side of the cone, shifted by $h$.

Next we study the effect of the polarization of the incident beam on the scattered light. Let the incident beam propagate along the $y$ axis (Fig. S2) and is polarized along the $x$ axis. The constructive x-ray interference will be observed from those parts of the sample, in which the helical axis is tilted by an angle $\pi/2 - \theta_{sc}/2$ with respect to the direction of the incident beam, where $\theta_{sc}$ is the scattering angle, i.e. for all the helices lying on the cone, as shown in Fig. S2. For the helix axis in the $yz$ plane, the incident beam is $\sigma$-polarized. For the helix axis in the $xy$ plane this incident beam is $\pi$-polarized and for the other directions on this cone the incident beam has both the $\sigma$ and $\pi$ component.

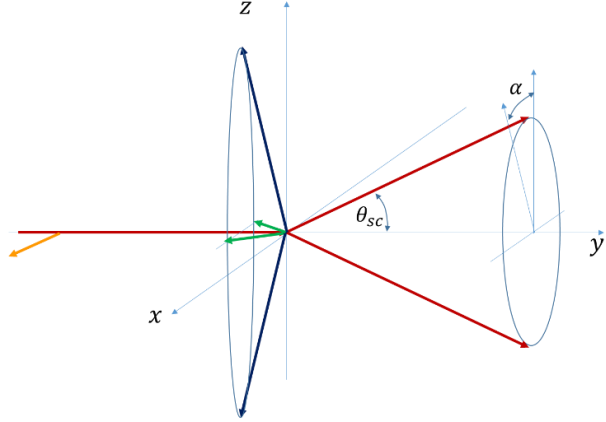

**Fig. S2.** Scattering geometry. The incident beam propagates along the $y$ axis and is polarized along the $x$ axis (orange arrow). The scattering is constructive for all the directions of the heliconical axes lying on the cone. For the directions of the helix axes denoted by blue arrows the incident beam is $\sigma$-polarized and for the "green" directions the beam is $\pi$-polarized. The scattering angle is $\theta_{sc}$ and $\alpha$ is the azimuthal angle.

For the beam scattered at the azimuthal angle $\alpha$ the $\sigma$ and $\pi$ components of the incident beam are:

$$\vec{\sigma}_{in} = (\cos\alpha, 0, 0) \ , \ \vec{\pi}_{in} = (0, 0, \sin\alpha)$$

and for the scattered beam the directions of the $\sigma$ and $\pi$ component are:

$$\vec{\sigma}_{sc} = (1, 0, 0) \ , \vec{\pi}_{sc} = (0, -\sin\theta_{sc}, \cos\theta_{sc}) \ .$$

The amplitudes of the $\sigma$ and $\pi$ components of the scattered light are found as [2]:

$$A_{\sigma\sigma} = \sum_{i,j} \sigma_{sc,j} F_{ij} \sigma_{in,i} \ ,$$

$$A_{\sigma\pi} = \sum_{i,j} \pi_{sc,j} F_{ij} \sigma_{in,i} \ ,$$

$$A_{\pi\sigma} = \sum_{i,j} \sigma_{sc,j} F_{ij} \pi_{in,i} \ ,$$

$$A_{\pi\pi} = \sum_{i,j} \pi_{sc,j} F_{ij} \pi_{in,i} \ .$$

where the elements of the tensors $F_{ij}$ are the elements of tensors $F_{q_0}^{(TB)}$, $F_{2q_0}^{(TB)}$ or $F_{2q_0}^{(SB)}$ rotated by an angle $\theta_{sc}/2$ around the $x$ axis for the helix axis to be in the proper direction with respect to the incident beam for the constructive interference to occur. The intensity of the scattered light is then calculated as [2]

$$I = |A_{\sigma\sigma}|^2 + |A_{\pi\sigma}|^2 + |A_{\sigma\pi}|^2 + |A_{\pi\pi}|^2 \ .$$

We find:

$$I_{q_0}^{(TB)} = \sin 2\theta_{TB} \cos^2\left(\frac{\theta_{sc}}{2}\right),$$

$$I_{2q_0}^{(TB)} = \frac{\left(-\frac{3}{f_{12}} + (2f_{12} + 1)\cos(2\theta_{TB})\right)^2}{4(2f_{12} + 1)^2}\left(1 + \sin^2\left(\frac{\theta_{sc}}{2}\right)\right) \times \left(\cos^2\alpha + \sin^2\left(\frac{\theta_{sc}}{2}\right)\sin^2\alpha\right),$$

$$I_{2q_0}^{(SB)} = 2(3 + \cos\theta_{sc})\left(\cos^2\alpha + \cos^2\left(\frac{\theta_{sc}}{2}\right)\sin^2\alpha\right),$$

where only the relevant factors were left required to compare the relative magnitudes of intensities. Note that the scattering angle depends on the magnitude of $q$. The value of the parameter $f_{12} = f_1/f_2$ determines the intensity of the $2q_0$ peak with respect to the intensity of the $q_0$ peak. If $f_{12} \geq 1$ the intensity of the $2q_0$ peak is approximately two orders of magnitude lower than the intensity of the $q_0$ peak. If $f_{12} < 1$, the intensity of the $2q_0$ increases rapidly. The intensity scattered from a powder sample as a function of the azimuthal angle on the screen for the incident beam along the $y$ direction and polarized along the $x$ axis, is given in Fig. S3.

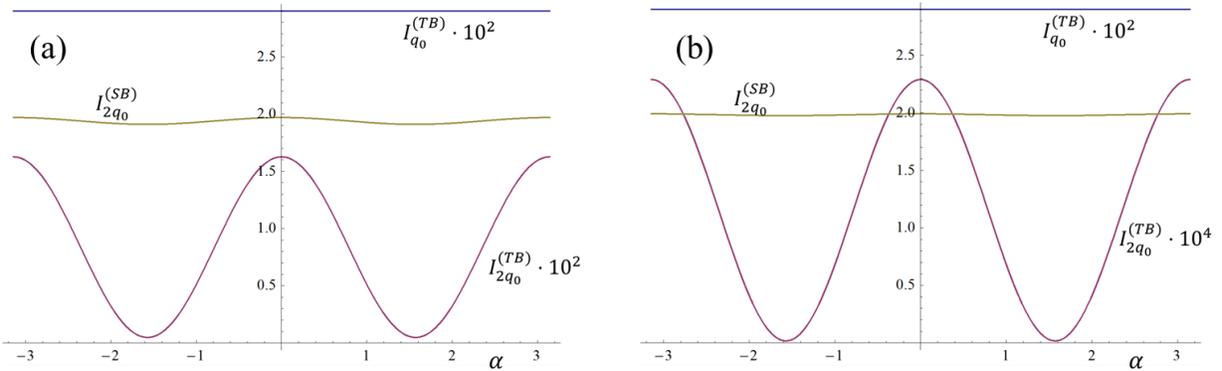

**Fig. S3.** The intensity ($I$) of the scattered beam (in arbitrary units) as a function of the azimuthal angle $\alpha$. The scattering geometry is shown in Fig. S2; (a) $f_{12} = 0.8$ and (b) $f_{12} = 1.0$. Parameter values: $\theta_{sc}(q_0) = 10°$, $\theta_{TB} = 10°$.

In experiment we observe a temperature dependence of the signal intensity. The intensity increases if temperature decreases and the material is deeper in the modulated nematic phase. The intensity of the $2q_0$ peak in the N$_{SB}$ phase is independent of the modulation angle $\theta_{SB}$, while the intensity of both the $q_0$ and $2q_0$ peaks in the N$_{TB}$ phase depend on the heliconical angle $\theta_{TB}$. Because the heliconical angle is temperature dependent [4] and the measured intensity is temperature dependent, we can discard the N$_{SB}$ structure as a possible structure of the modulated nematic phase. Experimentally, only one diffraction peak is observed with no clear dependence of the intensity on the direction of the polarisation of the incident beam. We can therefore conclude that the observed signal corresponds to the full pitch band ($q_0$) in the N$_{TB}$ modulated nematic phase. The half pitch band (the $2q_0$ peak) should be strongly polarization dependent. Although the intensity of this signal is expected to be much lower (assuming $f_{12} = 1$) than for the full pitch band (Fig. S4(a)), it should nevertheless be observed in the synchrotron measurements because the x-ray beam has a sufficient brightness. Moreover, as we will show later in the part of the Supporting Material related to the blue phase, we can expect that the parameter $f_{12}$ is lower than 1 and in that case the $2q_0$ signal intensity becomes comparable to the intensity of the $q_0$ peak (Fig. S4(b)).

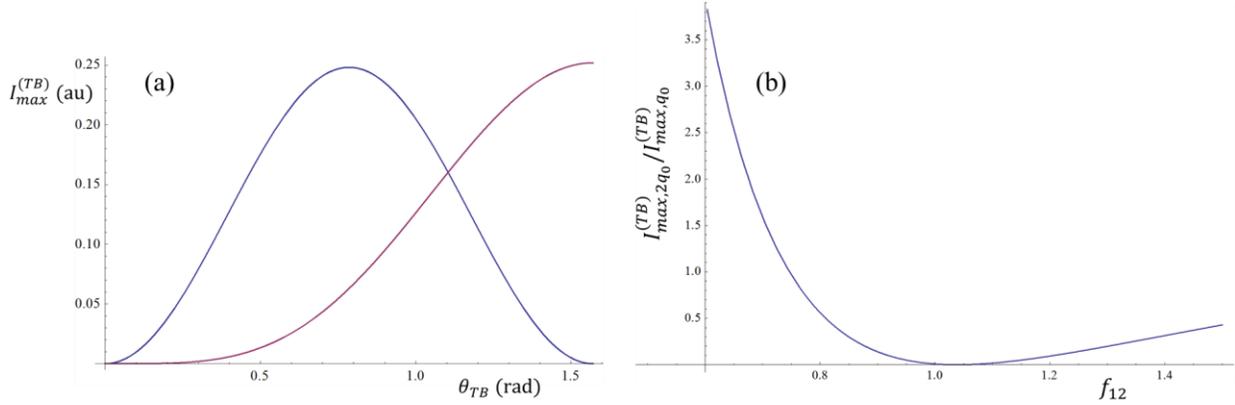

**Fig. S4.** (a) The maximum intensity ($I_{max}$) of the $q_0$ peak (blue) and the maximum intensity of the $2q_0$ peak (violet) in the N$_{TB}$ phase as a function of the heliconical angle $\theta_{TB}$. Parameter values: $f_{12} = 1$, $\theta_{sc}(q_0) = 10°$. (b) The ratio between the maximum intensity of the $2q_0$ and $q_0$ peaks in the N$_{TB}$ phase as a function of the parameter $f_{12}$ at $\theta_{TB} = 10°$, $\theta_{sc}(q_0) = 10°$ and $\theta_{sc}(2q_0) = 20°$.

The absence of the $2q_0$ peak in the diffraction pattern thus gives an important hint that the structure of the N$_{TB}$ is not of a simple heliconical type. Therefore, we considered a modified structure of the N$_{TB}$ structure, assuming that it is built of two interlocked helical modulations mutually shifted as shown in Fig. S1(c). The dispersion corrections to the form factor for such a structure are easy to obtain. We add the form factors for two interlocked helices, including the phase shift between them, noting that $F_{2q0}^{(TB)}$ is an even and $F_{q0}^{(TB)}$ is an odd function in $\theta_{TB}$, and obtain:

$$F_{2q_0} = F_{2q_0}^{(TB)}\left(1 + e^{i2q_0h}\right) ,$$
$$F_{q_0} = F_{q0}^{(TB)}\left(1 - e^{iq_0h}\right) .$$

If $q_0h \ll 1$, the dispersion correction to the form factor goes to zero for the $q_0$ peak, as expected from the symmetry consideration only (the pitch length reduces to $L/2$). If, however, the second helix is shifted by $L/4$, then the intensity of the $2q_0$ peak is zero. It should be pointed out, that the intensity of the $2q_0$ peak is zero if the helix is shifted by $L/4$ and the molecules are on the other side of the cone, as shown in Fig. S1, but is zero also for a pure shift by $L/4$ (with the molecules staying at the same side of the cone).

### Blue phase II:

The position of the non-resonant peaks in the blue phase II (BPII) is calculated in the following way. An infinite line of a uniform electron density along the cylinder axis (Fig. S5(a)) represents a cylinder. The lines of uniform density, as shown in Fig. S5(b), represent the defects. The scattering vector $\vec{q}$ is given by $\vec{q} = q_l(h, k, l)$, where $q_l = \frac{2\pi}{a}$ and $a$ is the size of the crystallographic unit cell (and equals half the pitch of the helix, see Fig. S5(a)) and $h, k$ and $l$ are the Miller indices.

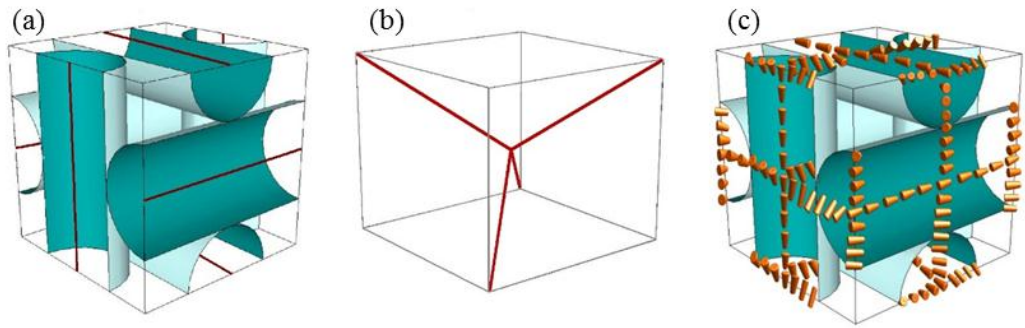

**Fig. S5.** Structure of the BPII phase. In the calculation of the form factor of the unit cell the cylinders (a) and defects (b) are represented by lines. (c) In the calculation of the tensorial correction to the form factor we consider spatial variation of the director in the direction perpendicular to the cylinder long axis.

The BPII phase consists of cylinders with their axes along the $x, y$ and $z$ direction. The Fourier transform of each line representing a cylinder gives a delta function, either $\delta(q_x)$, $\delta(q_y)$ or $\delta(q_z)$, where $q_i, i = x, y, z$, are the components of the scattering vector. Taking into account the position of the cylinders in the unit cell we find the following form factor ($F_c$) for the BPII phase:

$$F_c = \delta(q_x) + \delta(q_y)e^{i\pi(h+l)} + \delta(q_z)e^{i\pi k} \ .$$

From the expression for $F_c$ it follows, that at least one of the indices $h, k$ or $l$ has to be zero in order to have the form factor different from zero. If only one index is zero, the form factor is always different from zero. If two indices are zero, then the nonzero index has to be an even number to obtain a nonzero form factor; thus there is no $(0,0,1)$ peak, but there is a $(0,0,2)$ peak.

The defect lines run along $\frac{a}{2}(1,1,1) + \lambda(-1,1,1)$, $\frac{a}{2}(1,1,1) + \lambda(1,-1,1)$, $\frac{a}{2}(1,1,1) + \lambda(1,1,-1)$ and $\frac{a}{2}(1,1,1) - \lambda(1,1,1)$, where $\lambda \in \left[0, \frac{a}{2}\right]$. We assume, that along the defect lines the electron density differs from the electron density of the surrounding, so the form factor related to the defects ($F_d$) is given by the Fourier transform of the electron density along the defect lines. We find:

$$F_d = \frac{(1 - e^{-i(h+k-l)\pi})}{h+k-l} e^{2i(h+k)\pi} + \frac{(1 - e^{-i(h-k+l)\pi})}{h-k+l} e^{2i(h+l)\pi} + \frac{(1 - e^{-i(-h+k+l)\pi})}{-h+k+l} e^{2i(k+l)\pi}$$
$$- \frac{1 - e^{i(h+k+l)\pi}}{h+k+l} \quad .$$

The non-zero intensity peaks are given in Table S1. Note, that the defects do not give any $(0,0,l)$ peaks.

**Table S1.** Theoretically calculated peaks. Ticks denote the peaks for which a non-zero intensity is predicted. In the BPI, the (002) peak is a purely resonant peak and in BPII, the (001) peak is a resonant peak. The yellow and light green areas denote the peaks allowed by the $I4_132$ symmetry of the BPI phase and the $P4_232$ symmetry of the BPII phase, respectively. Note, that the (031) peak in the BPI is not resonantly enhanced. The peak (222) is not marked as purely resonant because it is allowed by the $I4_132$ symmetry; it is a coincidence that for the chosen model of the electron density (along the cylinder axis and the defect lines) the intensity in the elastic scattering is zero. The magnitude of the lattice wave vector is $q_l$ and the magnitude of the wave vector corresponding to the director modulation is $q_0$.

| | **BP-II** ($q_l = 2q_0$) | | | **BP-I** ($q_l = q_0$) | | |
|---|---|---|---|---|---|---|
| Peak ($hkl$) | Elastic due to cylinders | Elastic due to defects | Resonant/resonantly enhanced | Elastic due to cylinders | Elastic due to defects | Resonant/resonantly enhanced |
| **(001)** | ✗ | ✗ | ✓ | ✗ | ✗ | ✗ |
| **(011)** | ✓ | ✓ | ✓ | ✓ | ✗ | ✓ |
| **(111)** | ✗ | ✓ | ✓ | ✗ | ✗ | ✗ |
| **(002)** | ✓ | ✗ | ✓ | ✗ | ✗ | ✓ |
| **(021)** | ✓ | ✗ | ✓ | ✗ | ✗ | ✗ |
| **(211)** | ✗ | ✓ | ✓ | ✗ | ✓ | ✓ |
| **(022)** | ✓ | ✓ | ✓ | ✓ | ✓ | ✓ |
| **(221)** | ✗ | ✓ | ✓ | ✗ | ✗ | ✗ |
| **(031)** | ✓ | ✗ | ✓ | ✓ | ✗ | ✗ |
| **(311)** | ✗ | ✓ | ✓ | ✗ | ✗ | ✗ |
| **(222)** | ✗ | ✗ | ✗ | ✗ | ✗ | ✓ |
| **(032)** | ✓ | ✗ | ✓ | ✗ | ✗ | ✗ |
| **(321)** | ✗ | ✓ | ✓ | ✗ | ✓ | ✓ |

It is important to point out, that in the elastic x-ray scattering, although allowed by the symmetry, none of the diffraction signals will be observed, because the cylinders are in close contact (i.e. the density in the unit cell is almost uniform) and the density difference between the defect lines and cylinders, if any, is very low, because the phase has only a short range positional order. However, as shown below, the peaks observed in the resonant x-ray scattering are actually at the positions expected for the non-resonant peaks due to the 'resonant enhancement' effect [5, 6].

As in the case of the modulated nematic phase, to calculate the dispersion correction to the form factor, we consider the molecules having a rod-like shape. We present the calculations starting from the anisotropic traceless tensor $F_{ei}$ in the eigensystem, this time assuming that $f_1 = f_2$ in order to make the analytical results not too complicated. At the end we shall comment on the effect of $f_1 \neq f_2$.

With the $x$-axis along the average orientation of the long molecular axis, the anisotropic traceless tensor is:

$$F_{ei} = \begin{pmatrix} -2f_1 & 0 & 0 \\ 0 & f_1 & 0 \\ 0 & 0 & f_1 \end{pmatrix} .$$

Because the long molecular axis rotates, for example, along the $z$ axis (see Fig. S5(c)), the tensorial dispersion correction in the laboratory frame is obtained as

$$F = R^T F_{ei} R \quad ,$$

where $R$ is a rotation matrix:

$$R = \begin{pmatrix} \cos(q_0 z) & \sin(q_0 z) & 0 \\ -\sin(q_0 z) & \cos(q_0 z) & 0 \\ 0 & 0 & 1 \end{pmatrix} .$$

$q_0$ is the wave vector of the nematic director modulation and its magnitude is half the magnitude of the lattice wave vector $q_l$. We find (omitting all the irrelevant factors) that the dispersion correction to the form factor due to the helical rotation of the long molecular axis along the $z$ direction ($F_z$) is:

$$F_z = f_1 \begin{pmatrix} -\delta(q_z) - \dfrac{3}{2}\,\delta(q_z \pm q_l) & \pm\dfrac{3}{2}i\,\delta(q_z \pm q_l) & 0 \\ \pm\dfrac{3}{2}i\,\delta(q_z \pm q_l) & -\delta(q_z) + \dfrac{3}{2}\,\delta(q_z \pm q_l) & 0 \\ 0 & 0 & 2\delta(q_z) \end{pmatrix} .$$

The dispersion correction depends only on the $z$ component of the scattering vector and it is different from zero only for $l = 0$ or $l = 1$. Similarly, we find the dispersion correction due to the rotation along the $x$ ($F_x$) and $y$ direction ($F_y$). The dispersion correction to the form factor ($F_{hkl}$) of the unit cell is:

$$F_{hkl} = F_x\,(h)e^{i\pi k} + F_y(k)e^{i\pi l} + F_z(l)e^{i\pi h} \quad .$$

The full expression of $F_q$ is too comprehensive to be written in full, so we give some specific examples. First, we point out, that the resonant scattering should give one peak, which is forbidden in the elastic scattering, the $(0,0,1)$ peak. We also point out that those peaks, which have at least one of the Miller indices either 0 or 1, are resonantly enhanced, because this is the required condition for at least one dispersion correction to the form factor ($F_x$, $F_y$ or $F_z$) to be different from zero.

### *Blue phase I:*

We shall now repeat the above-described procedure to calculate the elastic and resonant peaks of the BPI phase. The schematic presentation of the BPI is shown in Fig. S6. To calculate the peaks allowed by the elastic x-ray scattering we again represent the structure of the double twist cylinders and defects as lines of a non-zero electron density and find:

$$F_c = \frac{1}{h}\left(-1 + e^{2ih\pi}\right)\left(e^{\frac{ik\pi}{2}} + e^{\frac{1}{2}i(3k+2l)\pi}\right) + \frac{1}{k}\left(-1 + e^{2ik\pi}\right)\left(e^{\frac{3il\pi}{2}} + e^{\frac{1}{2}i(2h+l)\pi}\right)$$
$$+ \frac{1}{l}\left(-1 + e^{2il\pi}\right)\left(e^{\frac{ih\pi}{2}} + e^{\frac{1}{2}i(3h+2k)\pi}\right)$$

and

$$F_d = -\frac{i\left(e^{ih\pi} - e^{i(-h+2(k+l))\pi}\right)}{2(h-k-l)\pi} - \frac{i\left(e^{2i(h+k)\pi} - e^{2il\pi}\right)}{2(h+k-l)\pi} + \frac{i\left(e^{i(2k+l)\pi} - e^{i(2h+3l)\pi}\right)}{2(h-k+l)\pi}$$
$$+ \frac{i\left(e^{ik\pi} - e^{i(k+2(h+k+l))\pi}\right)}{2(h+k+l)\pi} \quad .$$

The peaks allowed by the symmetry of the BPI phase are given in Table S1.

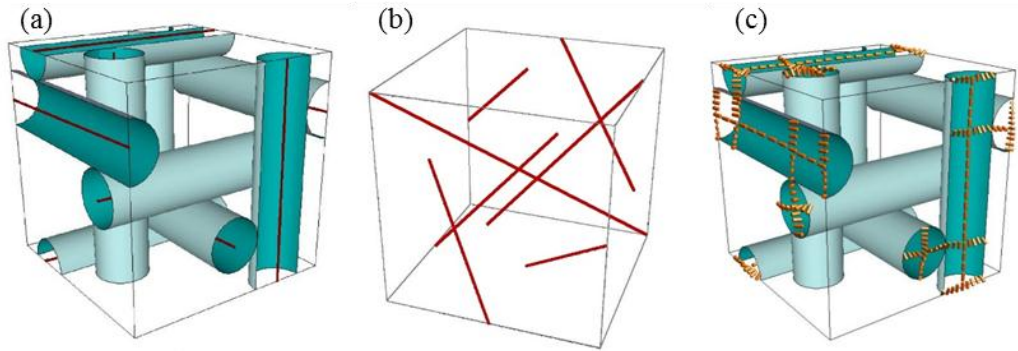

**Fig. S6.** Structure of the BPI phase. In the calculation of the form factor of the unit cell the (a) cylinders and (b) defects are represented by lines. (c) In the calculation of the tensorial correction to the form factor we consider spatial variation of the director in the direction perpendicular to the cylinder long axis.

To calculate the allowed resonant peaks we repeat the procedure used for the BPII phase, taking into account the positions of the helical double twist cylinders in the BPI phase. An excellent presentation of the structure in the BPI phase is available in the video clip [7]. There, it can be seen, that at each crossing of two cylinders there is a helical deformation of the orientation of long molecular axes. The long molecular axis rotates by $\pi$ when moving from the surface of one cylinder to the surface of the other one (see Fig.S 6(c)). So, the major difference with the BPII phase is, that now we do not have infinite helices, but several half-pitch helices in each unit cell.

To calculate the dispersion correction to the form factor of the unit cell there are a few more points to be considered. In calculating the traceless anisotropy tensor in the laboratory system for the helices along the $x$, $y$ and $z$ axis, we follow the procedure given for the BPII phase. If the long molecular axis rotates around the $z$-direction, the rotation angle ($\varphi_z$) is expressed as $\varphi_z = \pm\frac{\pi}{4} - q_0 z$ and in the integration (when calculating the Fourier transform) $z$ runs from 0 to $\pi/q_0$ (note, that in the BPI phase $q_0 = q_l$). By choosing the orientation of the molecules in one cylinder, we have defined the orientation of molecules in all cylinders. In addition, we choose, that the rotations along the $z$ axis are clockwise, and that the orientation of the long molecular axis is along the $x$ axis if $\varphi_z = 0$.

Similarly, for the helices along the $x$ axis, we set $\varphi_x = 0$ for the long molecular axis along the $y$ axis and then find that the rotation is clock-wise, thus $\varphi_x = \pm\frac{\pi}{4} - q_0 x$. For the helices along the $y$ axis, the long molecular axis is along the $z$ axis at $\varphi_y = 0$, the rotation is anticlock-wise and $\varphi_y = \pm\frac{\pi}{4} + q_0 y$.

There are 12 helices in one unit cell. For each one of them the initial $\varphi$, which is either $\pi/4$ or $-\pi/4$, has to be determined and then the phase factor due to the position of the helix has to be added to the proper form factor. Finally, we obtain the dispersion correction ($F_{hkl}$) to the form factor:

$$
\begin{aligned}
F_{hkl} = F_x^{(-)} &\left( e^{2i\pi(h/8 + 0/8k + 2/8l)} + e^{2i\pi(5h/8 + 4/8k + 6/8l)} \right) \\
+ F_x^{(+)} &\left( e^{2i\pi(-h/8 + 0/8k + 6/8l)} + e^{2i\pi(3h/8 + 4/8k + 2/8l)} \right) \\
+ F_z^{(+)} &\left( e^{2i\pi(4h/8 + 2/8k - 1/8l)} + e^{2i\pi(0h/8 + 6/8k + 3/8l)} \right) \\
+ F_z^{(-)} &\left( e^{2i\pi(4h/8 + 6/8k + 1/8l)} + e^{2i\pi(0h/8 + 2/8k + 5/8l)} \right) \\
+ F_y^{(+)} &\left( e^{2i\pi(6h/8 + 1/8k + 0/8l)} + e^{2i\pi(2h/8 + 5/8k + 4/8l)} \right) \\
+ F_y^{(-)} &\left( e^{2i\pi(6h/8 + 3/8k + 4/8l)} + e^{2i\pi(2h/8 - 1/8k + 0/8l)} \right) \quad ,
\end{aligned}
$$

where $F_{x,y,z}$ are the Fourier transforms of the helices along the $x$, $y$ and $z$ axis, respectively, and the index $+$ or $-$ denotes the helix, in which the initial $\varphi$ is $\pi/4$ or $-\pi/4$, respectively. The expression for $F_x$ is:

$$
F_x^{(\pm)} = \begin{pmatrix} F_{11}^{(\pm)} & 0 & 0 \\ 0 & F_{22}^{(\pm)} & F_{23}^{(\pm)} \\ 0 & F_{23}^{(\pm)} & F_{33}^{(\pm)} \end{pmatrix}
$$

with

$$
F_{11}^{(\pm)} = f_1 \frac{i\left(1 - e^{ih\pi}\right)}{h} \quad ,
$$

$$F_{22}^{(\pm)} = if_1 \frac{\left(1 - e^{ih\pi}\right)(-4 \pm 6ih + h^2)}{2h(4 - h^2)} \quad,$$

$$F_{23}^{(\pm)} = F_{23}^{(\pm)} = \pm 3if_1 \frac{\left(1 - e^{ih\pi}\right)h}{2(4 - h^2)} \quad,$$

$$F_{33}^{(\pm)} = if_1 \frac{\left(1 - e^{ih\pi}\right)(-4 \mp 6ih + h^2)}{2h(4 - h^2)} \quad.$$

Therefore, $F_x^{(\pm)}$ depends only on the Miller index $h$. The expressions for the other two tensors can be deduced from the above expressions: $h$ has to be replaced by $k$ or $l$ and the zeros are in the second or third line of the tensor.

In the BPI phase, the purely resonant peak (the peak not allowed by the $I4_132$ symmetry) is the (002) peak. The rest of the signals are allowed by the $I4_132$ symmetry, but are observed only in the resonant x-ray scattering due to the resonant enhancement: (011), (211), (022), (222), (321). However, not all the signals allowed by the $I4_132$ symmetry that are observed in the resonant experiment are resonantly enhanced, e.g. the (031) signal. The presence of a very week (031) signal in the RSoXS pattern might indicate that there is some distortion of the helical structure of the double twist cylinders.

Next, we study the effect of the polarization of the incident beam on the intensity of the scattered beam. As before, we choose the direction of the incident beam to be along the $y$ axis and is polarized along the $x$ axis. For the scattered light with the $\vec{q}$-vector lying in the $yz$ plane, the incident beam is $\sigma$-polarized, for $\vec{q}$ in the $xy$ plane it is $\pi$-polarized and for all the other direction both $\sigma$ and $\pi$ components are present. To obtain the amplitudes of the scattered light the procedure is more comprehensive than in the case of the modulated nematic phase. To show it, we consider the (002) peak as an example. The multiplicity of this peak is 6: $(00 \pm 2)$, $(0 \pm 20)$ and $(\pm 200)$. First, we calculate the dispersion correction of a given peak and for the positive values of the Miller index 2 we find:

$$F_{002} = f_1 \begin{pmatrix} -3 & i & 0 \\ i & 3 & 0 \\ 0 & 0 & 0 \end{pmatrix}, F_{020} = f_1 \begin{pmatrix} 3 & 0 & i \\ 0 & 0 & 0 \\ i & 0 & -3 \end{pmatrix}, F_{200} = f_1 \begin{pmatrix} 0 & 0 & 0 \\ 0 & -3 & i \\ 0 & i & 3 \end{pmatrix}.$$

To have the $\vec{q}$ vector along the $x, y$ or $z$ axis, one should choose a proper direction of the incident beam. Since we have fixed the direction of the incident beam, we have to find those crystals in the powder sample, that will have their local $x, y$ or $z$ axis (for which the above expressions of $F_{hkl}$ apply) at a proper angle with respect to the direction of the incoming beam. The directions of these $\vec{q}(hkl)$ in the laboratory system are defined by angles $\theta_0$ and $\varphi_0$, defining the angle between the laboratory $z$ axis and $\vec{q}(hkl)$ and the $y$ axis and $\vec{q}(hkl)$, respectively. For $\vec{q}(002)$ both angles are zero, for $\vec{q}(020)$ $\varphi_0 = 0$ and $\theta_0 = \pi/2$ and for $\vec{q}(200)$ $\varphi_0 = \pi/2$ and $\theta_0 = \pi/2$. We thus have to rotate the $F_{hkl}$ tensor by

$$F_{hkl}' = R_{\theta_0}^T \left(R_{\varphi_0}^T F_{hkl} R_{\varphi_0}\right) R_{\theta_0} \quad,$$

where

$$R_{\varphi_0} = \begin{pmatrix} \cos\varphi_0 & \sin\varphi_0 & 0 \\ -\sin\varphi_0 & \cos\varphi_0 & 0 \\ 0 & 0 & 1 \end{pmatrix}$$

and

$$R_{\theta_0} = \begin{pmatrix} 1 & 0 & 0 \\ 0 & \cos\left(\dfrac{\theta_{sc}}{2} + \theta_0\right) & \sin\left(\dfrac{\theta_{sc}}{2} + \theta_0\right) \\ 0 & -\sin\left(\dfrac{\theta_{sc}}{2} + \theta_0\right) & \cos\left(\dfrac{\theta_{sc}}{2} + \theta_0\right) \end{pmatrix} .$$

As in the case of modulated nematics, for the beam scattered at the azimuthal angle $\alpha$ (see Fig. S2), the $\sigma$ and $\pi$ components of the incident beam are:

$$\vec{\sigma}_{in} = (\cos\alpha, 0, 0) ,$$
$$\vec{\pi}_{in} = (0, 0, \sin\alpha) .$$

For the scattered beam the directions of the $\sigma$ and $\pi$ component are:

$$\vec{\sigma}_{sc} = (1, 0, 0) ,$$
$$\vec{\pi}_{sc} = (0, -\sin\theta_{sc}, \cos\theta_{sc}) ,$$

where the magnitude of the scattering angle depends on the chosen peak. The intensity of the scattered peak is now calculated following the procedure given in previous section. In the case of the (002) peak there are 6 contributions to the intensity of this peak and they are all the same. In general, these contributions can be different. For example, in the case of the (112) peak, which has a multiplicity of 24, the contribution of the (112) peak is different from the contributions of the (211) and (121) peaks.

Figure S7 gives the intensity of the peaks (011), (002) and (112), i.e. the peaks with the lowest magnitudes of $q$, as a function of the azimuthal angle at $f_{12} = 1$ (the ratio used in the above presented calculation) and $f_{12} = 0.7$. It can be seen that the polarization dependence of the (112) peak depends on the value of $f_{12}$. If $f_{12} < 0.8$, the (112) peak has the same azimuthal intensity dependence as the peaks (011) and (002), which is in agreement with the experimental observations (see Fig. 3 in main text). The integrated intensity ($I^{(int)}$) of the (002) peak is higher than for the (011) peak, but they are still of the same order of magnitude: at $f_{12} = 1$, the ratio is $I_{002}^{(int)}/I_{011}^{(int)} = 1.4$, at $f_{12} = 0.7$, the ratio is $I_{002}^{(int)}/I_{011}^{(int)} = 1.6$. Experimentally, the integrated intensity of the (011) peak is higher than $I^{(int)}$ of the (002) peak, but still of the same order of magnitude. Because of the crudeness of the model one can expect only qualitative agreement, so we do not find this discrepancy disturbing. The integrated intensity of the (112) peak is the lowest, which agrees with experimental observations.

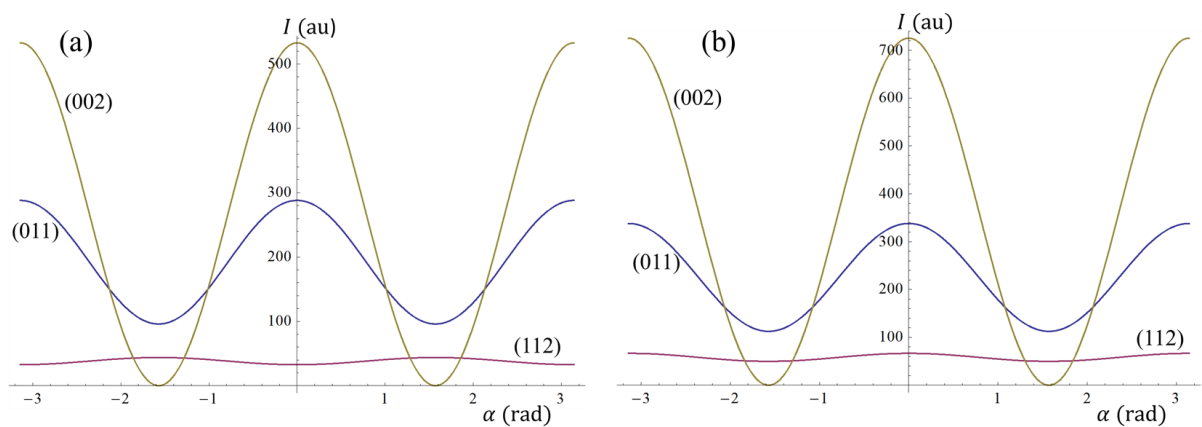

**Fig. S7.** The intensity ($I$) in arbitrary units (au) as a function of the azimuthal angle for the peaks $(011)$, $(002)$ and $(112)$, which are the peaks with the lowest magnitudes of $q$; (a) $f_{12} = 1$, (b) $f_{12} = 0.7$.

## 2. *Materials and methods*

### *Materials:*

All the studied compounds are dimers with an odd number of atoms in the linkage between the mesogenic cores, which induces a bent molecular geometry. Two of them, **AZO7** and **SB3**, are built from asymmetric molecules bearing a chiral, cholesteric unit, the third one, **CB7CB**, is a symmetric dimer (Fig. S8). The materials were synthesised at the University of Warsaw, the synthesis, purification and characterization was described previously [8].

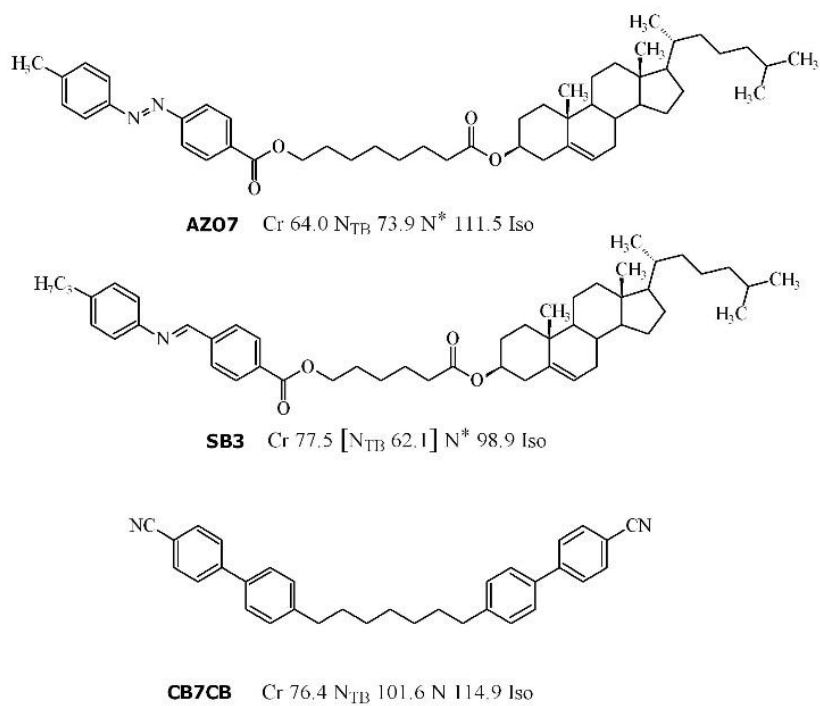

**AZO7**    Cr 64.0 $N_{TB}$ 73.9 N* 111.5 Iso

**SB3**    Cr 77.5 [$N_{TB}$ 62.1] N* 98.9 Iso

**CB7CB**    Cr 76.4 $N_{TB}$ 101.6 N 114.9 Iso

**Fig. S8.** Molecular structure of the studied compounds, **AZO7**, **SB3** and **CB7CB**. For each compound a phase sequence and phase transition temperatures determined by the differential scanning calorimetry (DSC) are given. Note, that for **AZO7** a narrow range of a blue phase between the N* and Iso has been found by microscopic observations; however, it was not recorded on the DSC curves due to a limited resolution. Upon cooling the **AZO7** samples, the blue phase was metastable down to the transition to the NTB phase and thus the cholesteric phase was not observed.

### *Methods:*

The x-ray experiments were performed on the soft x-ray scattering beam line (11.0.1.2) at the Advanced Light Source of Lawrence Berkeley National Laboratory. The x-ray beam was tuned to the K-edge of carbon absorption, ~280 eV (Fig. S9).

The x-ray beam with a cross-section of 300 × 200 μm was linearly polarized, with the polarization direction that can be continuously changed from the horizontal to vertical. Samples with thickness lower than $1\,\mu m$ were placed between two 100-nm-thick $Si_3N_4$ slides. The scattering intensity was recorded using the Princeton PI-MTE CCD detector, cooled to $-45°C$, having a pixel size of $27\,\mu m$, with an adjustable distance from the sample. The detector was translated off axis to enable a recording of the diffracted x-ray intensity. The adjustable position of the detector allowed to cover a broad range of $q$ vectors, corresponding to periodicities from approximately $5.0 - 300$ nm.

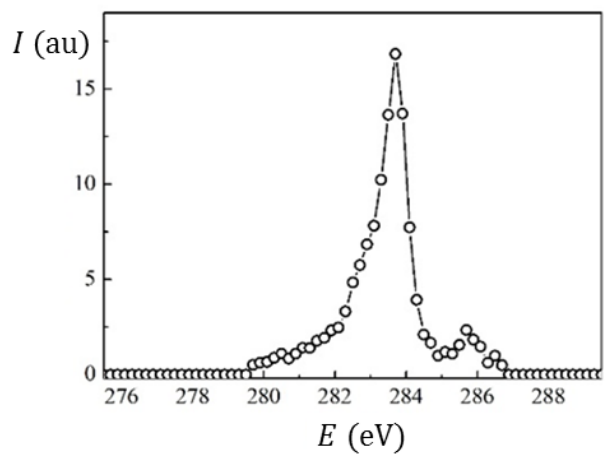

**Fig. S9.** Intensity ($I$) of the signal in the $N_{TB}$ phase as a function of the energy ($E$) of the x-ray beam, measured for AZO7 compound.

The AFM images were taken with the Bruker Dimension Icon microscope, working in the tapping mode at the liquid crystalline-air surface. Cantilevers with a low spring constant, $k = 0.4$ N/m were used, the resonant frequency was in a range of $70 - 80$ kHz, a typical scan frequency was 1 Hz. Samples for the AFM imaging were prepared in glass cells at elevated temperature, quenched to room temperature and unsealed.

### 3. *Additional experimental results*

In both, the $N_{TB}$ and N* phase of chiral dimer **SB3** the resonant diffraction signal shows only a weak temperature dependence (Fig. S10).

For the achiral dimer **CB7CB** the heliconical structure changes critically when temperature approaches the transition to the non-modulated nematic phase, while in the crystalline phase the resonant signal related to the helical structure was found to be temperature independent (Fig. S11). For the **AZO7** material a direct transition from the BPI to $N_{TB}$ phase was observed on cooling, with a small temperature range in which both phases coexists (Fig. S12).

The AFM image of **SB3** recorded at the ambient temperature shows that an additional structure with a longer periodicity ($50 - 80$ nm) is present in the $N_{TB}$ phase (Fig. S13).

The AFM images taken at a room temperature of the crystalline and metastable $N_{TB}$ phase of **CB7CB** are shown in Fig. S14. Both phases are characterized by $\sim 8$ nm periodicities, however in the case of a crystalline phase no focal conics were found. The fast Fourier transform (FFT) obtained from the image registered in the crystalline phase shows the first and second harmonic, and both are of almost equally intensity. On the other hand, the FFT obtained from the image registered in the $N_{TB}$ phase shows only the first harmonic.

The position of the RSoXS signal in the $N_{TB}$ phase of the **AZO7** compound is temperature dependent (Fig. S15).

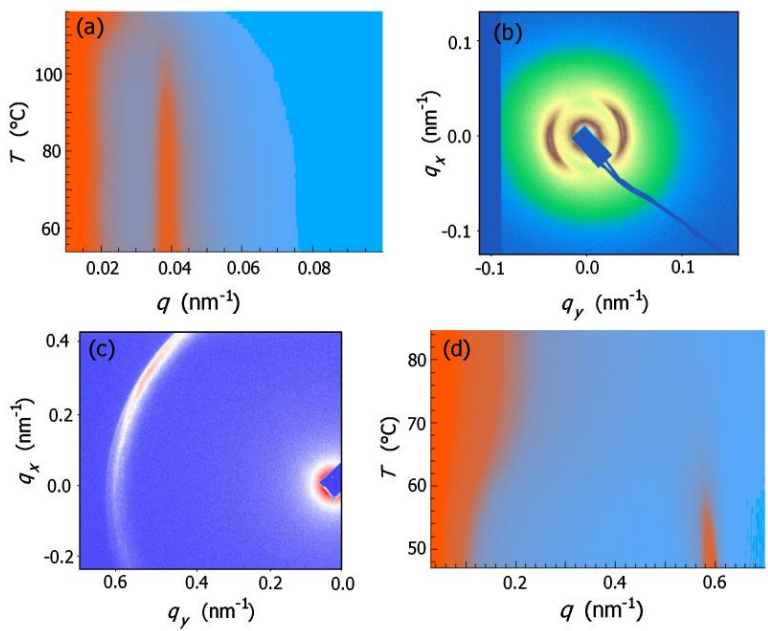

**Fig. S10.** Temperature dependence of the RSoXS pattern measured for **SB3** compound in the temperature range of the (a) N* and (d) $N_{TB}$ phase. The patterns were obtained in the consecutive cooling/heating runs with modified experimental conditions (different detector position). (b) and (c) give the 2D RSoXS patterns taken in the N* and $N_{TB}$ phase, respectively.

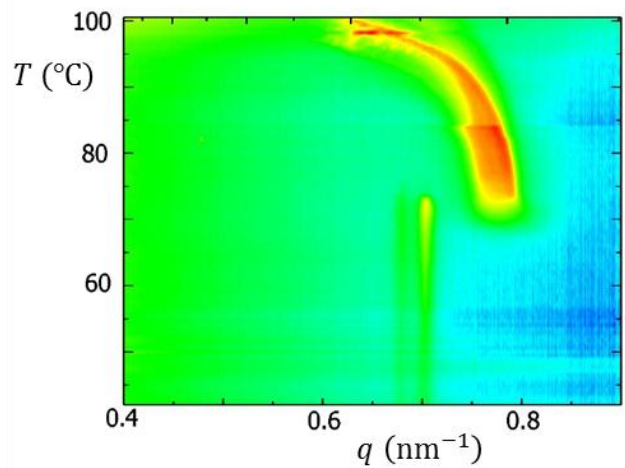

**Fig. S11.** Temperature evolution of the RSoXS signal for the **CB7CB** sample, measured on heating. Note that while in the crystalline phase the signal position is practically temperature independent, in the $N_{TB}$ phase it changes strongly due to the changes of the heliconical pitch, from 8 to 10 nm upon approaching the transition to the N phase.

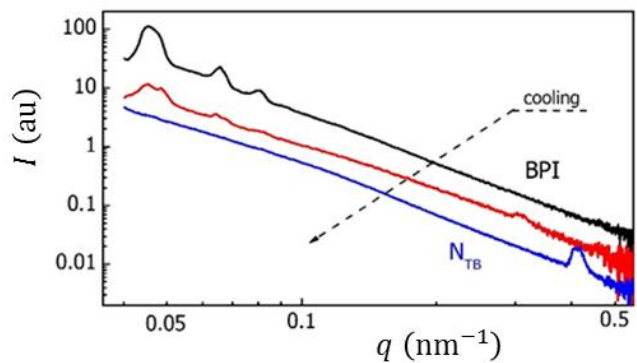

**Fig. S12.** RSoXS patterns, scattered intensity ($I$) in arbitrary units (au) as a function of the magnitude of the scattering vector $q$, for the **AZO7** compound recorded in a broad $q$ range. On cooling, a direct transition from the BPI (black line) to the $N_{TB}$ phase (blue line) is observed with a small (1 K) temperature range of the phase coexistence (red line).

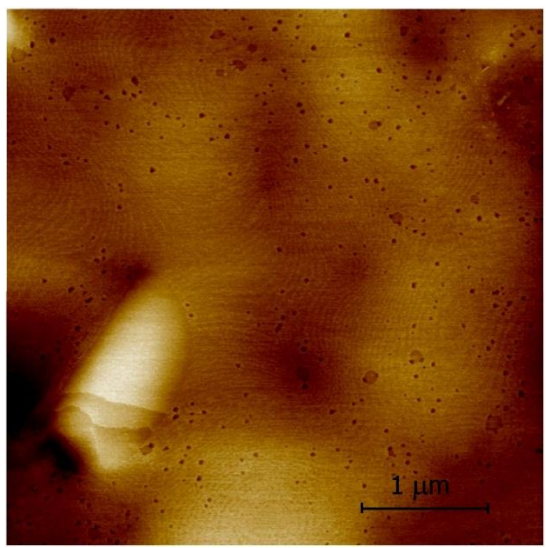

**Fig. S13.** AFM image taken in the $N_{TB}$ phase of **SB3** compound at room temperature. The distance between the lines is approximately 50-80 nm.

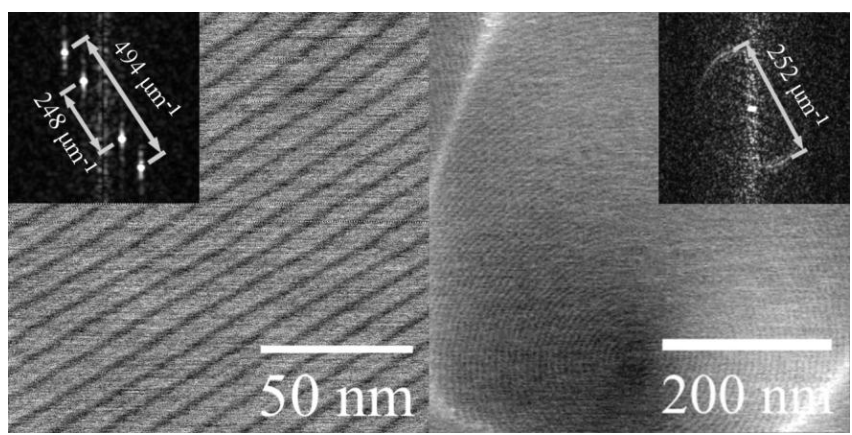

**Fig. S14.** AFM images of **CB7CB** in a crystal (left) and $N_{TB}$ (right) phase and the corresponding fast Fourier transform patterns. In the crystalline phase the intensity of the first and second harmonic of the detected periodicity (8.06 nm) is of the same order of magnitude. In the $N_{TB}$ phase only a weak first harmonic (7.93 nm) is detected.

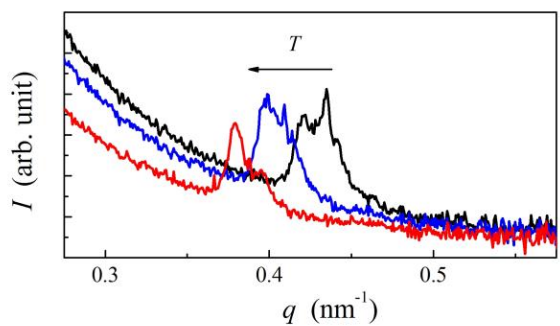

**Fig. S15:** RSoXS patterns, the intensity ($I$) in arbitrary units as a function of the magnitude of the scattering vector $q$, for the **AZO7** compound in the $N_{TB}$ phase at three different temperatures.

## REFERENCES

[1] D. H. Templeton and L. K. Templeton, Acta Crystallogr. Sect. A **36**, 237, (1980).

[2] V. E. Dmitrienko, Acta Crystallogr. Sect. A **39**, 29, (1983).

[3] A. M. Levelut and B. Pansu, Phys. Rev. E **60**, 6803, (1999).

[4] C. Meyer, G. R. Luckhurst, and I. Dozov, J. Mater. Chem. C **3**, 318, (2015).

[5] D. H. Templeton and L. K. Templeton, Acta Crystallogr. Sect. A **38**, 62, (1982).

[6] V. E. Dmitrienko, Acta Crystallogr. Sect. A **40**, 89, (1984).

[7] H. Yoshida, https://www.youtube.com/watch?v=NJG14FdcWkg

[8] E. Gorecka, N. Vaupotic, A. Zep, D. Pociecha, J. Yoshioka, J. Yamamoto and H. Takezoe, Angew. Chem. Int. Ed. **54**, 10155, (2015).